\documentclass[12pt]{article}
\usepackage{color}
\usepackage{amssymb,latexsym,bm, amsmath,amsthm, amsfonts,multirow, graphicx,mathrsfs}
\usepackage{eufrak,booktabs, ulem}
\usepackage{epsfig}
\usepackage{enumitem}
\usepackage{float}
\graphicspath{{fig/}}

\def\Vb{{\boldsymbol V}}

\def\Xb{{\boldsymbol X}}

\def\vec{{\rm vec}}

\def\1{{\bm 1}}
\def\0{{\bm 0}}

\def\rmspan{{\rm span}}

\def\one{{(1)}}

\def\IFs{{\rm IF2}}

\def\Sr{\mathcal{S}_r}

\def\Sr{\mathcal{S}_r}
\newcommand{\real}{\mathbb{R}}

\newtheorem{prop}{Proposition}
\newtheorem{thm}{Theorem}

\newtheorem{lem}[thm]{Lemma}

\newtheorem{rmk}{Remark}
\newtheorem{cor}[thm]{Corollary}

\def\boxit#1{\vbox{\hrule\hbox{\vrule\kern6pt\vbox{\kern6pt#1\kern6pt}\kern6pt\vrule}\hrule}}

\begin{document}

%\title{The $\phi$-PCA framework: efficiency-loss-free PCA and its ordering-robust harmonic mean variant}
\title{The $\phi$-PCA Framework: A Unified and Efficiency-Preserving Approach with Robust Variants}

\author{Hung Hung$^1$, Zhi-Yu Jou$^2$, Su-Yun Huang$^2$, and Shinto Eguchi$^3$ \\[0ex]
\small $^1$Institute of Health Data Analytics and Statistics,
National
Taiwan University, Taiwan\\
\small $^2$Institute of Statistical Science, Academia Sinica, Taiwan\\
\small $^3$Institute of Statistical Mathematics, Japan
}

\date{}
\maketitle

\begin{abstract}

Principal component analysis (PCA) is a fundamental tool in multivariate statistics, yet its sensitivity to outliers and limitations in distributed environments restrict its effectiveness in modern large-scale applications. To address these challenges, we introduce the $\phi$-PCA framework which provides a unified formulation of robust and distributed PCA. The class of $\phi$-PCA methods retains the asymptotic efficiency of standard PCA, while aggregating multiple local estimates using a proper $\phi$ function enhances ordering-robustness, leading to more accurate eigensubspace estimation under contamination. Notably, the harmonic mean PCA (HM-PCA), corresponding to the choice $\phi(u)=u^{-1}$, achieves optimal ordering-robustness and is recommended for practical use. Theoretical results further show that robustness increases with the number of partitions, a phenomenon seldom explored in the literature on robust or distributed PCA. 
Altogether, the partition-aggregation principle underlying $\phi$-PCA offers a general strategy for developing robust and efficiency-preserving methodologies applicable to both robust and distributed data analysis.

\noindent \textbf{Key words:} aggregation; distributed computing; efficiency; ordering-robustness; PCA; random partition.
\end{abstract}

\clearpage

\section{Introduction}\label{sec.introduction}

Principal component analysis (PCA) is one of the most widely used tools in multivariate statistics and data science, serving as the default method for dimension reduction and feature extraction (Jolliffe, 2002). However, standard PCA is fragile in modern high-dimensional data environments, where two challenges are particularly acute: robustness and distributed computing. 
Robustness is critical because even a small fraction of contaminated data can severely distort the estimated principal components. To overcome the non-robustness issue of standard PCA, classical robust PCA methods (Li and Chen, 1985; Campbell, 1980; Croux and Haesbroeck, 2000; Maronna, 2005; Hubert et al., 2005) mainly developed down-weighting strategies to mitigate the influence of outliers, though some loss of efficiency is often inevitable under clean data. Distributed computing is equally important, as contemporary applications often involve datasets spread across multiple or federated sites, where centralization is computationally or administratively infeasible. To this end, distributed PCA methods have been developed to address scalability by combining local eigen-decompositions or subspace-related statistics from multiple sites to form a unified estimate. See Wu et al. (2018) for a comprehensive review, Fan et al. (2019) for aggregation via projection matrices, and Jou et al. (2025) for weighted aggregation using eigenvalue information. This line of work represents an important step toward scalable eigen-decomposition, although robustness considerations remain limited within this framework.

While the two challenges above may seem unrelated, recent studies have shown that the core idea of distributed PCA (i.e., aggregating multiple results to form a unified estimate) provides a new perspective for improving the robustness of PCA. Specifically, Hung and Huang (2025), by extending the data-partitioning idea of Yata and Aoshima (2010), proposed product PCA (PPCA) as a robust variant of standard PCA. PPCA randomly divides the data into two partitions, yielding two subsample covariance matrices $\widehat{S}_1$ and $\widehat{S}_2$, which are then aggregated via the geometric mean $\widehat{S}_1^{\frac 12}\widehat{S}_2^{\frac 12}$ to estimate the leading eigenvectors for dimension reduction. Hung and Huang (2025) show that, in the presence of outliers, PPCA yields a more accurate ordering of the signal eigenvectors than standard PCA, thereby improving ordering-robustness. The robustness gain of PPCA stems from the fact that an outlier is unlikely to contaminate both subsample covariance matrices simultaneously; thus, the combination of partitioning and product aggregation enhances stability. Notably, in the absence of outliers, PPCA and standard PCA are shown to share the same asymptotic distribution, demonstrating that PPCA achieves ordering-robustness while maintaining efficiency. Consequently, PPCA provides a practical and theoretically appealing enhancement to PCA for real-world applications.

Considering the potential merits of PPCA, this paper generalizes and examines the effects of its two distinctive features: random partitioning of the data and product-based aggregation. We propose the $\phi$-PCA framework, which extends PPCA by allowing an arbitrary number of partitions $m$ and a flexible family of aggregation rules indexed by a function $\phi$. This partition–aggregation principle unifies standard PCA, robust PCA, and distributed PCA within a single formulation. We further establish that all members of the $\phi$-PCA family preserve the asymptotic efficiency of standard PCA, while suitable choices of $\phi$ yield additional ordering-robustness relative to standard PCA. A particularly notable case is the harmonic-mean PCA (HM-PCA), corresponding to $\phi(u)=u^{-1}$, which achieves optimal ordering-robustness among arithmetic-mean, geometric-mean, and harmonic-mean aggregation rules.

$\phi$-PCA differs from existing works in several important aspects.
First, unlike classical robust PCA methods that achieve robustness through down-weighting schemes, $\phi$-PCA achieves ordering-robustness purely via the partition-aggregation principle. Consequently, $\phi$-PCA lies outside the class of traditional robust approaches. It shares the same asymptotic distribution as standard PCA under clean data, whereas classical robust methods inevitably incur efficiency loss. Second, although both $\phi$-PCA and distributed PCA involve aggregating $m$ PCA results, the role of $m$ fundamentally differs. In distributed PCA, $m$ is typically fixed by the number of local sites, whereas in $\phi$-PCA it serves as a design parameter that can be tuned to control robustness. To the best of our knowledge, the role of $m$ as a robustness-controlling parameter has not been systematically explored in the literature, and our theoretical results show that robustness increases with the number of partitions. Moreover, the effect of $\phi$ on the robustness issue has received little attention in the distributed PCA literature.
While Jou et al. (2025) analyzed the stability of distributed PCA under a simple diagonal covariance contamination model, we provide a more general investigation of how the choice of $\phi$ affects ordering-robustness. Since distributed PCA can be viewed as a special case of $\phi$-PCA with a fixed $m$ and a specific $\phi$, our framework also extends and contributes to the distributed PCA literature. In particular, the optimal HM-PCA is especially relevant in applications where ordering-robustness is of primary importance.

 In summary, $\phi$-PCA offers a unified theoretical framework connecting robust PCA and distributed PCA. Our main contributions are as follows:
\begin{itemize}
\item[(i)]
We formalize the partition-aggregation principle and establish that all $\phi$-PCA methods preserve the asymptotic efficiency of standard PCA.

\item[(ii)]
We identify HM-PCA as achieving optimal ordering-robustness among the arithmetic-mean, geometric-mean, and harmonic-mean aggregation rules.

\item[(iii)]
We demonstrate, through theoretical analysis and numerical study, that $\phi$-PCA is a flexible and effective family of PCA methods.
\end{itemize}

The rest of the paper is organized as follows. 
Section~\ref{sec.phi_pca} introduces the class of efficiency-preserving $\phi$-PCA methods. Sections~\ref{sec.HM_PCA}-\ref{sec.compare} develop their statistical properties and establish the optimal ordering-robustness of HM-PCA. Section~\ref{sec.simulation} reports simulation studies evaluating performance, followed by a real-data application in Section~\ref{sec.data}. Concluding remarks are given in Section~\ref{sec.conclusion}, and all proofs are collected in the Appendix.

\section{The $\bm\phi$-PCA Framework}\label{sec.phi_pca}

Let $X$ be a $p$-dimensional random vector with distribution $F$, mean vector $\mu$, and positive definite covariance matrix $\Sigma$. For simplicity and without loss of generality, we assume $\mu=0$. The spectral decomposition of $\Sigma$ is
$\Sigma = \Gamma \Lambda \Gamma^\top$,
where $\Lambda = \mathrm{diag}(\lambda_1,\ldots,\lambda_p)$ consists of $p$ distinct eigenvalues in descending order, and $\Gamma = [\gamma_1,\ldots,\gamma_p]$ is the corresponding orthonormal matrix of eigenvectors. Let $r$ be the target rank with $\lambda_r \gg \lambda_{r+1}$, so that the leading $r$-dimensional eigensubspace 
\[
\Sr = \mathrm{span}(\Gamma_r) \quad{\rm with}\quad \Gamma_r = [\gamma_1,\ldots,\gamma_r]
\]
is well separated from the remaining tail eigenvectors. A central goal of PCA is to estimate $\Sr$, thereby enabling subsequent analyses to be performed on the projection of $X$ onto this subspace. Note that the ordering of eigenvectors is determined by their corresponding eigenvalues, and hence the terms ``ordering of eigenvectors'' and ``ordering of eigenvalues'' will be used interchangeably hereafter. This also highlights an important message that the accurate recovery of $\Sr$ requires not only correct estimation of the eigenvectors but also preservation of the proper ordering of the eigenvalues, as any misordering can bias the estimation of $\Sr$.

The proposed $\phi$-PCA framework generalizes PPCA by allowing for an arbitrary number of random partitions and a flexible aggregation method. Let $\Xb = [X_1,\ldots,X_n]^\top$ be the $n\times p$ data matrix of a random sample $\{X_i\}_{i=1}^n$ from $F$. For a given $m\in \mathbb{N}$, $\phi$-PCA starts by constructing the size-$m$ random partition of $\Xb$ as
\begin{eqnarray}
[\Xb^{(m)}]=\{\Xb_1,\ldots,\Xb_m\},\label{random_partition}
\end{eqnarray}
where each $\Xb_k=[X_{k,1},\ldots,X_{k,\frac{n}{m}}]^\top$ is of the size $\frac{n}{m}\times p$ (assume $\frac{n}{m}$ to be an integer for notation simplicity). Let $\widehat S_k$ be the sample covariance matrix of the subsample $\Xb_k$, and
\begin{eqnarray}
[\widehat S^{(m)}]=\{\widehat S_1,\ldots,\widehat S_m\}.
\end{eqnarray}
To aggregate $[\widehat S^{(m)}]$, let $\phi(\cdot):\real \to \real^+$ be a positive and strictly monotone function. Consider the following transformation induced by $\phi$, which is defined on the set of nonnegative definite matrices with spectral decomposition $\Sigma=\sum_{j=1}^{p}\lambda_j\gamma_j\gamma_j^\top$:
\begin{eqnarray}
\phi(\Sigma) &=& \sum_{j=1}^p\phi(\lambda_j)\gamma_j\gamma_j^\top.
\end{eqnarray}
We propose to aggregate $[\widehat S^{(m)}]$ via the {\it $\phi$-generalized mean} of $[\widehat S^{(m)}]$ defined as
\begin{eqnarray}\label{M_phi}
\widehat\Sigma_\phi^{(m)}=\phi^{-1}\left(\frac{1}{m}\sum_{k=1}^m\phi(\widehat S_k)\right).
\end{eqnarray}
Due to the random partition mechanism, each $\widehat S_k$ serves as an estimator of $\Sigma$, and the definition of $\phi(\cdot)$ ensures that the eigenvectors of $\widehat\Sigma_\phi^{(m)}$ are likewise estimators of eigenvectors of $\Sigma$. This implies that conducting PCA based on $\widehat\Sigma_\phi^{(m)}$, denoted by $\phi$-PCA, is a valid PCA procedure. Consider the spectral decomposition
\[\widehat\Sigma_\phi^{(m)}=\widehat\Gamma\widehat\Lambda\widehat\Gamma^\top\quad{\rm with}\quad \widehat{\Gamma} = [\widehat{\gamma}_1,\ldots,\widehat{\gamma_p}]\quad{\rm and}\quad\widehat{\Lambda} = \mathrm{diag}(\widehat{\lambda}_1,\ldots,\widehat{\lambda}_p).\]
$\phi$-PCA estimates $\Sr$ by $[\widehat\gamma_1,\ldots,\widehat\gamma_r]$, where the ordering of $\widehat\gamma_j$ is determined by the magnitude of $\widehat\lambda_j$. The following theorem establishes the asymptotic properties of $\phi$-PCA.

%Standard PCA proceeds by computing the sample covariance matrix $\widehat{S}$ of $\Xb$, followed by its spectral %decomposition $\widehat{S} = \widetilde{\Gamma}\widetilde{\Lambda}\widetilde{\Gamma}^\top$, 
%where $\widetilde{\Lambda} = \mathrm{diag}(\widetilde{\lambda}_1,\ldots,\widetilde{\lambda}_p)$ and $\widetilde{\Gamma} = %[\widetilde{\gamma}_1,\ldots,\widetilde{\gamma_p}]$. Standard PCA then estimates $\Sr$ by %$\mathrm{span}(\widetilde{\gamma}_1,\ldots,\widetilde{\gamma}_r)$, where the ordering of $\widetilde\gamma_j$ is %determined by the magnitude of $\widetilde\lambda_j$. 

\begin{thm}[Efficiency-preserving asymptotic normality of $\phi$-PCA]\label{thm.asymptotic_normality}
Assume that $\phi$ is positive, strictly monotone and has a continuous second derivative. Let $\xi=({\rm diag}(\Lambda)^\top,\vec(\Gamma)^\top)^\top$ and $\widehat\xi_\phi^{(m)}=({\rm diag}(\widehat\Lambda)^\top,\vec(\widehat\Gamma)^\top)^\top$. Then, for fixed $m$ and $p$, as $n\to\infty$, we have the weak convergence
\begin{eqnarray*}
\sqrt{n}\left(\widehat\xi_\phi^{(m)}-\xi\right)&\stackrel{d}{\to}& N(\0_{p(p+1)},\Vb),
\end{eqnarray*}
where $\Vb=H^\top{\rm cov}\{\vec(XX^\top)\}H$ with $H=[\gamma_1\otimes\gamma_1,\ldots,\gamma_p\otimes\gamma_p, \gamma_1\otimes M_1,\ldots,\gamma_p\otimes M_p]$ and $M_j=(\lambda_jI_p-\Sigma)^+$.
\end{thm}

A key point of Theorem~\ref{thm.asymptotic_normality} is that $\Vb$ depends only on $(\Lambda, \Gamma)$ and is invariant to the choice of $(m, \phi)$. This indicates that $\phi$-PCA and standard PCA share the same asymptotic distribution for eigenvalues and eigenvectors, noting that standard PCA corresponds to the special case $m = 1$ (see Remark~\ref{rmk.phi_pca_m1}). This invariance has two important implications:
\begin{itemize}
\item
In the absence of outlier, using $\phi$-PCA to estimate $(\Lambda,\Gamma)$ incurs no loss of efficiency relative to standard PCA, for any $(m,\phi)$. 
\item
This invariance effectively decouples efficiency from robustness, allowing $(m,\phi)$ to be treated as design choices. It further enables us to study the effects of $m$ and $\phi$, and to optimize the ordering-robustness of $\phi$-PCA without compromising efficiency.
\end{itemize}

\begin{rmk}\label{rmk.phi_pca_m1}
The function $\phi$ plays a role in $\phi$-PCA only when $m>1$, since $\widehat\Sigma_\phi^{(1)} = \widehat S$. In other words, standard PCA is a special case of $\phi$-PCA with $m = 1$.
\end{rmk}

\section{The Ordering-Robustness of $\bm\phi$-PCA}\label{sec.HM_PCA}

The aim of this section is to investigate the ordering-robustness of $\phi$-PCA, with focus on a subfamily of power $\phi$-PCA methods. A specific choice $\phi(u)=u^{-1}$ is then identified to possess the optimal ordering-robustness property, which is called Harmonic Mean-PCA (HM-PCA).

%We begin by analyzing the general behavior of $\phi$-PCA under data perturbations. We then turn our attention to a %subfamily of power $\phi$-PCA methods defined by $\phi(u) = u^\beta$, with particular focus on the cases $\beta = 1$, %$\beta \to 0$, and $\beta = -1$, corresponding to arithmetic, geometric, and harmonic mean aggregation, respectively.

\subsection{Perturbation and stability analysis of $\bm\phi$-PCA}\label{sec.hm_pca}

To investigate the ordering-robustness of $\phi$-PCA, we aim to examine how the $\phi$-PCA statistical functional is perturbed when the target population is contaminated by an outlier $x$. Since $\phi$-PCA involves a size-$m$ random partition $[\Xb^{(m)}]$,  its underlying perturbation mechanism differs from that of standard PCA. For standard PCA, the target population of $\Xb$ is $F$. To study its perturbation, it is natural to consider the contaminated distribution 
\[F_{x,\varepsilon}=(1-\varepsilon)F+\varepsilon\delta_x,\]
where $\varepsilon>0$ is small and $\delta_x$ denotes the Dirac measure concentrated at $x$. For $\phi$-PCA, the target population of $[\Xb^{(m)}]$ is $F^{(m)}:=(F_1,\ldots,F_m)=(F,\ldots,F)$, where the last equality follows from the random partition mechanism. Since an occasionally occurred outlier can hardly affect more than one subsample of $[\Xb^{(m)}]$, 
we consider the following perturbation for $\phi$-PCA:
\begin{eqnarray}\label{perturbation}
F^{(m)}\to F_{x,\varepsilon}^{(m)} := (F_{x,m\varepsilon}, F,\ldots,F).
\end{eqnarray}
The factor $m\varepsilon$ arises because each subdata matrix $\Xb_k$ contains only $\frac{n}{m}$ samples, thereby amplifying the relative influence of the outlier $x$ by a factor of $m$. Let $\theta$ denote a parameter of interest. We define $\theta(F_{x,\varepsilon}^{(m)})$ to be the $\phi$-PCA statistical functional associated with a given pair $(m,\phi)$ evaluated at $F_{x,\varepsilon}^{(m)}$. Standard PCA is recovered as a special case with $m=1$ (see Remark~\ref{rmk.phi_pca_m1}), in which case $\theta(F_{x,\varepsilon}^{(1)})$ coincides with the statistical functional of standard PCA. Note that the $\phi$-PCA statistical functional is invariant to which component of $F^{(m)}$ receives the contamination. Hence, without loss of generality, we assume in (\ref{perturbation}) that the first component of $F^{(m)}$ is perturbed to $F_{x,m\varepsilon}$.

The following theorem characterizes the stability differences between the $\phi$-PCA and standard PCA statistical functionals when the underlying distribution is perturbed by the inclusion of a point $x$.

\begin{thm}[Perturbation of $\phi$-PCA vs.~standard PCA]\label{thm.IF2_phi_PCA}
Given $(m,\phi)$ and under the perturbation~(\ref{perturbation}), the following results hold for small $\varepsilon$:
\begin{eqnarray*}
\gamma_j^\top\left\{\gamma_j(F_{x,\varepsilon}^{(m)})-\gamma_j(F_{x,\varepsilon}^{(1)})\right\}&=&o(\varepsilon^2),\\
\sum_{j\ne k\le r}\gamma_k^\top\left\{\gamma_j(F_{x,\varepsilon}^{(m)})-\gamma_j(F_{x,\varepsilon}^{(1)})\right\}&=&\varepsilon^2(m-1)\sum_{j<k\le r}\frac{-(\lambda_jd_j+\lambda_kd_k)}{(\lambda_j-\lambda_k)^2}(\gamma_j^\top x)(\gamma_k^\top x) + o(\varepsilon^2),\\
\frac{1}{\lambda_j}\left\{\lambda_j(F_{x,\varepsilon}^{(m)})-\lambda_j(F_{x,\varepsilon}^{(1)})\right\}&=&\varepsilon^2
(m-1)\left\{d_j\left(x^\top W_j x-1-\frac{\phi_j''\lambda_j}{\phi_j'}\right)+\frac{\phi_j''\lambda_j}{2\phi_j'}\right\}+ o(\varepsilon^2),
\end{eqnarray*}
where $d_j=(\gamma_j^\top x)^2/\lambda_j$, $W_j=\frac{\phi_j''}{2\phi_j'}P_j+M_j-\frac{1}{\phi_j'}\left\{\phi_jI_p-\phi(\Sigma)\right\}M_j^2$, $M_j=(\lambda_jI_p-\Sigma)^+$, $P_j=\gamma_j\gamma_j^\top$, $\phi_j=\phi(\lambda_j)$, $\phi_j'=\frac{d}{d\lambda_j}\phi(\lambda_j)$, and $\phi_j''=\frac{d^2}{d\lambda_j^2}\phi(\lambda_j)$.
\end{thm}

Theorem~\ref{thm.IF2_phi_PCA} shows that, up to terms of order $\varepsilon^2$, the cosine similarity between the perturbed eigenvector $\gamma_j(F_{x,\varepsilon}^{(m)})$ and $\gamma_j$ in $\phi$-PCA coincides with that of $\gamma_j(F_{x,\varepsilon}^{(1)})$ in standard PCA. Notably, this directional stability is independent of both the aggregation method $\phi$ and the partition size $m$. It explains why all members of the $\phi$-PCA family share the same asymptotic normality and are therefore free from efficiency loss, in agreement with Theorem~\ref{thm.asymptotic_normality}. 
Moreover, the total cross cosine similarity between $\{\gamma_j(F_{x,\varepsilon}^{(m)})-\gamma_j(F_{x,\varepsilon}^{(1)})\}$ and $\gamma_k$, summed over all $(j,k)$ with $j\ne k\le r$, is also invariant to the choice of $\phi$. This suggests that the class of $\phi$-PCA methods is relatively robust and tends to exhibit similar perturbation patterns in the presence of outliers when estimating the signal eigenvectors $\{\gamma_j\}_{j\le r}$. In contrast, the eigenvalue behavior differs: the choice of $\phi$ can affect the second-order perturbation of $\lambda_j(F_{x,\varepsilon}^{(m)})$ which plays a key role in determining the ordering of $\gamma_j(F_{x,\varepsilon}^{(m)})$. 
In summary, Theorem~\ref{thm.IF2_phi_PCA} indicates that the differences in performance among the $\phi$-PCA methods in estimating $\mathcal{S}_r$ mainly arise from their ability to preserve the correct ordering of $\{\gamma_j(F_{x,\varepsilon}^{(m)})\}_{j\le r}$ (see Remark~\ref{rmk.ordering} for further discussion). It is therefore of interest to identify the $\phi$
function that best achieves this task, a question we examine in detail in the next subsection.

\begin{rmk}\label{rmk.ordering}
Consider the most severe case where the outlier point $x$ lies in the orthogonal complement of the signal subspace, i.e., $x\in \Sr^\perp$, a reasonable working assumption when $p\gg r$. In this setting, such an $x$ cannot affect the signal eigenvectors $\{\gamma_j\}_{j\le r}$ and eigenvalues $\{\lambda_j\}_{j\le r}$, but can alter the magnitudes of the noise eigenvalues $\{\lambda_k\}_{k>r}$, thereby increasing the risk of mis-ordering the signal components. Specifically, when $x\in\Sr^\perp$, Theorem~\ref{thm.IF2_phi_PCA} simplifies to $\gamma_j(F_{x,\varepsilon}^{(m)})=0$ and $\lambda_j(F_{x,\varepsilon}^{(m)})=0$ for any $j\le r$ and $(m,\phi)$, while $x$ and $(m,\phi)$ still influence the perturbation 
$\frac{1}{\lambda_k}\{\lambda_k(F_{x,\varepsilon}^{(m)})-\lambda_k(F_{x,\varepsilon}^{(1)})\}$, $k>r$, which, in turn, can affect the relative ordering of all eigenvectors $\gamma_j(F_{x,\varepsilon}^{(m)})$'s. This highlights the importance of maintaining the correct ordering of the signal eigenvectors when estimating $\Sr$.
\end{rmk}

\subsection{The optimal ordering-robustness of HM-PCA} \label{sec.hm_pca.opt}

To study the impact of the function $\phi(\cdot)$ on the ordering-robustness of $\phi$-PCA, we focus for the remainder of the discussion on the subclass of power functions
\begin{equation}\label{power_phi}
\phi(u)=u^\beta, \quad \beta \in \mathbb{R}\setminus \{0\}
\end{equation}
with the limiting case $\beta \to 0$ treated separately.
This choice yields a family of $\phi$-PCA methods indexed by $(m,\beta)$, where the aggregated covariance matrix is given by
\begin{eqnarray}\label{M_phi.beta}
\widehat\Sigma_\beta^{(m)}=\left(\frac{1}{m}\sum_{k=1}^m\widehat S_k^\beta\right)^{\frac{1}{\beta}}.
\end{eqnarray}
Three notable cases arise:
\begin{itemize}
\item
$\beta = -1$: aggregation via matrix harmonic-mean (HM),
\item
$\beta \to 0$: aggregation via matrix geometric-mean (GM),
\item
$\beta = 1$: aggregation via matrix arithmetic-mean (AM).
\end{itemize}

\begin{rmk}[Geometric mean case]\label{rmk.gm}
Any affine transformation of $\phi(\cdot)$, i.e., $a \phi(\cdot) + b$ with $a \in \mathbb{R} \setminus \{0\}$ and $b \in \mathbb{R}$, yields the same eigenvalues and eigenvectors in $\phi$-PCA. Thus, the limiting case $\beta\to 0$ can be defined using $\lim_{\beta\to 0}\frac{u^\beta-1}{\beta}=\ln(u)$. In this situation, the aggregated covariance matrix of $\phi$-PCA becomes $\exp\left\{\frac{1}{m}\sum_{k=1}^{m}\ln(\widehat{S}_k)\right\}$, which is a version of GM for $[\widehat S^{(m)}]$ proposed by Arsigny {\it et al.} (2007).
\end{rmk}

To quantitatively assess the ordering-robustness of $\phi$-PCA based on $\widehat\Sigma_\beta^{(m)}$, we adopt the same measure $\rho_{jk}$ introduced in Hung and Huang (2025) defined as
\begin{eqnarray}\label{eq.rho}
\rho_{jk} =\frac{\lambda_j}{\lambda_j+\lambda_k}\in(0.5,1),\quad j\le r < k.
\end{eqnarray}
Note that $\rho_{jk}$ quantifies the dominance of the signal eigenvalue $\lambda_j$ relative to the combined magnitude of $\lambda_j$ and the noise eigenvalue $\lambda_k$. A higher value of $\rho_{jk}$ indicates an increased chance that the estimated $\lambda_j$ will be correctly ranked ahead of the estimated $\lambda_k$. 
Then, under the perturbation~(\ref{perturbation}), define the total improvement of $\phi$-PCA over standard PCA as
\begin{eqnarray}\label{tau}
\tau(x)=\frac{1}{r(p-r)}\sum_{j\le r}\sum_{k>r}\frac{1}{\eta_{jk}}\Big\{\rho_{jk}(F_{x,\varepsilon}^{(m)})-\rho_{jk}(F_{x,\varepsilon}^{(1)})\Big\},
\end{eqnarray}
where $\eta_{jk}=\rho_{jk}(1-\rho_{jk})$. The scaling factor $\eta_{jk}^{-1}$ is included to make the differences $\{\rho_{jk}(F_{x,\varepsilon}^{(m)})-\rho_{jk}(F_{x,\varepsilon}^{(1)}):j\le r< k\}$ comparable over different values of $\rho_{jk}$. A positive value of $\tau(x)$ indicates that estimating the ordering of $\{\gamma_j\}_{j\le r}$ by $\phi$-PCA tends to be more robust to the influence of $x$ compared to standard PCA. We have the following results regarding the dependence of $\tau(x)$ on $(m,\beta)$.

%The performance of $\phi$-PCA in estimating $\Sr$  can therefore be examined by assessing its robustness in estimating %$\{\theta_j\}_{j\le r}$, where
%\begin{eqnarray}
%\theta_j = (\gamma_j^\top,\rho_{j,r+1},\ldots,\rho_{jp})^\top.\label{theta_j}
%\end{eqnarray}
%More specifically, by treating $\theta_j(\cdot)$ as a statistical functional (i.e., a mapping from the underlying data %distribution to the parameter of interest) we can compare the robustness of $\phi$-PCA and standard PCA under %distributional contamination, by evaluating their sensitivity to the inclusion of an outlying observation $x$.

\begin{thm}[Asymptotic expansion of ordering-robustness for $\phi$-PCA]\label{thm.tau.phi_PCA}
Assume the perturbation model~(\ref{perturbation}), and let $\phi (u)=u^\beta$. For $\beta \in \mathbb{R}\setminus\{0\}$ and small $\varepsilon$, the total improvement in ordering-robustness, as defined in~\eqref{tau}, is expressed as
\begin{eqnarray*}
\tau(x) &=& \varepsilon^2(m-1)\left\{\left(\frac{1-\beta}{2}x^\top\Sigma^{-1}x+ \beta\right)\Delta(x)+\Delta_\beta(x)\right\}+o(\varepsilon^2),
\end{eqnarray*}
where $\Delta(x)=\frac{\sum_{k>r}
d_k}{p-r}-\frac{\sum_{j\le r}d_j}{r}$ with $d_j=(\gamma_j^\top x)^2/\lambda_j$, and $\Delta_\beta(x)=\frac{\sum_{k>r} \xi_kd_k}{p-r}-\frac{\sum_{j\le r} \xi_jd_j}{r}$ with $\xi_j=\sum_{\ell\ne j}\left\{\frac{\lambda_\ell\lambda_j}{(\lambda_j-\lambda_\ell)^2}
\frac{1-(\lambda_\ell/\lambda_j)^\beta}{\beta}-\frac{\lambda_\ell}{\lambda_j-\lambda_\ell}-\frac{1-\beta}{2}\right\}d_\ell$.
\end{thm}

Theorem~\ref{thm.tau.phi_PCA} explicitly quantifies how the partition size $m$ and the aggregation parameter $\beta$ contribute to the robustness gain $\tau(x)$, thereby disentangling their respective roles in enhancing the ordering-robustness of $\phi$-PCA. As shown by Hung and Huang (2025), 
the term $\Delta(x)$, which is independent of $(m,\beta)$, tends to be positive in many practical settings. Since the leading term in $\tau(x)$  is proportional to $(m-1) \left( \frac{1-\beta}{2} x^\top \Sigma^{-1} x + \beta \right) \Delta(x)$, such positivity suggests that ordering-robustness is enhanced whenever $m>1$ and $\beta \le 1$. 
The effect of the remainder term $\Delta_\beta(x)$ is more subtle, as it depends jointly on $x$, $\beta$, and the eigenvalues $\{\lambda_j\}_{j=1}^p$. 
In particular, negative values of $\Delta_\beta(x)$ can reduce the overall magnitude of $\tau(x)$ and thus dampen robustness gains. To gain further insight into the ordering-robustness of $\phi$-PCA, we next examine the extreme case where $x\in \Sr^\perp$ (see Remark~\ref{rmk.ordering} for details), which leads to the following results.

\begin{cor}\label{cor.delta_beta} 
Assume the conditions of Theorem~\ref{thm.tau.phi_PCA}, and suppose that $x\in \Sr^\perp$. Then, for $\beta \in \mathbb{R} \setminus \{0\}$, we have
\begin{eqnarray*}
\Delta_\beta(x)=\frac{1}{p-r}\sum_{k>\ell>r}\left\{\beta-\frac{1}{\beta}\cdot\frac{(\frac{\lambda_\ell}{\lambda_k})^\beta+
(\frac{\lambda_k}{\lambda_\ell})^\beta-2}
{\frac{\lambda_\ell}{\lambda_k}+\frac{\lambda_k}{\lambda_\ell}-2}\right\}d_kd_\ell.
\end{eqnarray*}
Moreover, $\Delta_\beta(x)=0$ if and only if $\beta\in\{-1,0,1\}$, where the case $\beta=0$ is understood in the limiting sense $\beta\to 0$. In these cases $\beta\in\{-1,0,1\}$, the total improvement is
\begin{eqnarray*}
\tau(x) &=& \frac{\varepsilon^2(m-1)}{p-r}\left\{\left(\frac{1-\beta}{2}x^\top\Sigma^{-1}x+\beta\right)
\left(x^\top\Sigma^{-1}x\right)\right\}+o(\varepsilon^2)
\end{eqnarray*}
for small $\varepsilon$.
\end{cor}

Corollary~\ref{cor.delta_beta} shows that, when the outlier $x$ lies in $\Sr^\perp$, the ordering-robustness gain ($\tau(x) > 0$) of $\phi$-PCA is guaranteed for $\beta \in \{0,1\}$ whenever $m > 1$.
For $\beta=-1$, a gain is also expected if $x^\top\Sigma^{-1}x>1$, a condition that holds for most $x$ in high-dimensional settings (for example, $X^\top\Sigma^{-1}X\sim \chi^2_p$ when $X\sim N(0,\Sigma)$). Interestingly, the choices $\beta\in\{-1,0,1\}$ with vanishing $\Delta_\beta(x)$ correspond to HM-, GM- and AM-aggregation of $\phi$-PCA, respectively. The next theorem summarizes the total improvement for HM-, GM- and AM-PCA.

\begin{thm}[Ordering-robustness gains of HM-, GM-, and AM-PCA]\label{thm.robustness.3}
Assume the perturbation model~(\ref{perturbation}), and let $x\in \mathbb{R}^p$ be an arbitrary outlier point. For  small $\varepsilon$, we have the following results.
\begin{enumerate}
\item
The total improvement of HM-PCA is
\begin{eqnarray*}
\tau_{\rm HM}(x) &=& \varepsilon^2(m-1)\Big\{\left(x^\top \Sigma^{-1}x-1\right)\Delta(x)\Big\}+o(\varepsilon^2).
\end{eqnarray*}
If $x\in\Sr^\perp$, then $\tau_{\rm HM}(x) = \frac{\varepsilon^2(m-1)}{p-r}\left(x^\top \Sigma^{-1}x-1\right)\left(x^\top \Sigma^{-1}x\right)+o(\varepsilon^2)$, which guarantees the ordering-robustness gain of HM-PCA when $x^\top \Sigma^{-1}x>1$ and $m>1$.
\item 
The total improvement of GM-PCA is
\begin{eqnarray*}
\tau_{\rm GM}(x) &=& \varepsilon^2(m-1)\left\{\frac{1}{2}\left(x^\top \Sigma^{-1}x\right)\Delta(x)+\Delta_0(x)\right\}+o(\varepsilon^2).
\end{eqnarray*}
where $\Delta_0(x)=\frac{p}{r(p-r)}\sum_{j\le r}\sum_{k>r}\frac{\lambda_j\lambda_k}{(\lambda_j-\lambda_k)^2}\left\{\frac{1}{2}\left(\frac{\lambda_j}{\lambda_k}
-\frac{\lambda_k}{\lambda_j}\right)-\ln\frac{\lambda_j}{\lambda_k}\right\}d_jd_k\ge 0$.
If $x\in\Sr^\perp$, then $\tau_{\rm GM}(x) = \frac{\varepsilon^2(m-1)}{2(p-r)}\left(x^\top \Sigma^{-1}x\right)^2+o(\varepsilon^2)$, which guarantees the ordering-robustness gain of GM-PCA when $m>1$.

\item 
The total improvement of AM-PCA is
\begin{eqnarray*}
\tau_{\rm AM}(x) &=& \varepsilon^2(m-1)\Delta(x)+o(\varepsilon^2).
\end{eqnarray*}
If $x\in\Sr^\perp$, then $\tau_{\rm AM}(x) = \frac{\varepsilon^2(m-1)}{p-r}\left(x^\top \Sigma^{-1}x\right)+o(\varepsilon^2)$, which guarantees the ordering-robustness gain of AM-PCA when $m>1$.
\end{enumerate}
\end{thm}

%The implication of Theorem~\ref{thm.robustness.3} is twofold. First, random partition with a large $m$ has the potential %to enhance the ordering-robustness of $\phi$-PCA (see Section~\ref{sec.robust.beta1} for further discussion on this %matter). 

It is worth emphasizing that, although Corollary~\ref{cor.delta_beta} shows $\Delta_\beta(x)=0$ for $x \in \Sr^\perp$ if and only if $\beta \in \{-1,0,1\}$, the remainder term in fact satisfies $\Delta_{-1}(x)=\Delta_1(x)=0$ and $\Delta_0(x)\ge 0$ for general $x$. Consequently, Theorem~\ref{thm.robustness.3} shows that the ordering-robustness of these special cases is determined entirely by $m$ and $\Delta(x)$. 
The implications are twofold. 
\begin{itemize}
\item
Since $\Delta(x)$ tends to be positive  in many practical settings (Hung and Huang, 2025), using a larger random partition size $m$ generally yields greater ordering-robustness than standard PCA (see Section~\ref{sec.m} for the selection of $m$).
\item
The gain in ordering-robustness is more pronounced when $x$ is an influential outlier, i.e., when it has a large Mahalanobis distance  $(x^\top\Sigma^{-1}x)^{\frac12}$ with $\mu=0$. 
\end{itemize}
  
An important message from Theorem~\ref{thm.robustness.3} is that the ordering-robustness gains of HM-PCA, GM-PCA, and AM-PCA differ, with the coefficient $\frac{1-\beta}{2}$ in the term $x^\top\Sigma^{-1}x$ playing a key role. Since $\beta$ does not affect the asymptotic normality of $\phi$-PCA (Theorem~\ref{thm.asymptotic_normality}), we conclude that HM-PCA is optimal among $\beta\in\{-1,0,1\}$ for any $x\in\Sr^\perp$ satisfying $x^\top\Sigma^{-1}x>2$, as it maximizes the ordering-robustness gain among these cases. Combined with Theorem~\ref{thm.asymptotic_normality}, this yields the following result for HM-PCA. \\

\hrule
\vspace{0.25cm}
\centerline{\bf Optimal Ordering-Robustness of HM-PCA without Efficiency Loss}
\vspace{0.25cm}
\hrule

\begin{enumerate}[label=(\Alph*)]
\item\label{free_robustnessA}
In the absence of outliers, HM-PCA and standard PCA share the same asymptotic distribution for estimating $(\Lambda, \Gamma)$. In other words, HM-PCA incurs no efficiency loss relative to standard PCA when estimating $\mathcal{S}_r$ (see Theorem~\ref{thm.asymptotic_normality}).

\item\label{free_robustnessB}
In the presence of outliers, HM-PCA behaves similarly to standard PCA in terms of eigenvector perturbations for estimating $\{\gamma_j\}_{j\le r}$, but demonstrates greater robustness in preserving their ordering whenever an outlier $x$ with $\tau_{\rm HM}(x)>0$ occurs. Furthermore, for any $x \in \mathcal{S}_r^\perp$ satisfying $x^\top\Sigma^{-1}x>2$, HM-PCA achieves the largest total improvement among the three methods, AM, GM, and HM (see Theorem~\ref{thm.robustness.3}).
\end{enumerate}
\hrule
\vspace{0.7cm}

\noindent
This property supports the applicability of HM-PCA, as it suggests that HM-PCA generally does not perform worse than standard PCA in estimating $\Sr$, regardless of whether outliers are present.
Theorem~\ref{thm.robustness.3} also implies that GM-PCA, like HM-PCA, is ordering-robust without efficiency loss relative to standard PCA, although $\tau_{\rm GM}(x)$ and $\tau_{\rm HM}(x)$ differ in form. 
Likewise, PPCA has been shown to share this efficiency-loss-free property (Hung and Huang, 2025), but it lies outside the $\phi$-PCA family, with its total improvement derived from a different formulation.
These differences suggest that HM-PCA, GM-PCA, and PPCA may exhibit distinct ordering-robustness behaviors, and a detailed comparison is given in Section~\ref{sec.compare}.

We close this section by presenting the implementation algorithm.\\[-1ex]

\hrule
\vspace{0.25cm}
\centerline{\bf Implementation Algorithm for HM-PCA and GM-PCA}
\vspace{0.25cm}
\hrule

\begin{enumerate}
\item
Randomly partition the data matrix $\Xb$ into $m$ subsets $[\Xb^{(m)}]$, and compute the corresponding subsample covariance matrices $[\widehat S^{(m)}]=[\widehat S_1, \dots, \widehat S_m]$.

\item
Compute the aggregated covariance matrices
\[
\widehat\Sigma_{\rm HM}^{(m)} = \Bigg\{\frac{1}{m}\sum_{k=1}^m (\widehat S_k + \epsilon I_p)^{-1}\Bigg\}^{-1},
\qquad
\widehat\Sigma_{\rm GM}^{(m)} = \exp\Bigg\{\frac{1}{m}\sum_{k=1}^m \ln(\widehat S_k + \epsilon I_p)\Bigg\},
\]
where $\epsilon>0$ is a small regularization constant added to ensure numerical stability in the case of singular or nearly singular subsample covariance matrices.

\item
Obtain the eigenvalues and eigenvectors $(\widehat\Lambda, \widehat\Gamma)$ from the spectral decomposition of $\widehat\Sigma_{\rm HM}^{(m)}$ or $\widehat\Sigma_{\rm GM}^{(m)}$.
\end{enumerate}

\hrule
\vspace{3ex}

\noindent In Step~1, if $n$ cannot be divided by $m$, we randomly allocate the remaining samples to different partitions so that the number of samples contained in each $\Xb_k$ is either $\lfloor\frac{n}{m}\rfloor$ or $\lfloor\frac{n}{m}\rfloor + 1$. This modification will not affect the theoretical results. Note that $\widehat S_k$ is singular when $\frac{n}{m}\le p$. To avoid the matrix singularity, in Step~2 we add a ridge regularization term
$\epsilon I_p$ to $\widehat S_k$. This adjustment helps stabilize the HM and GM aggregation while preserving the relative insignificance of the tail eigenvectors. We recommend the choice $\epsilon=10^{-8}\{{\rm tr}(\widehat S)/p\}$, which keeps $\epsilon$ small relative to the average diagonal elements of $\widehat S$. This value is adopted in all our numerical studies.

\subsection{Determination of the partition size $\bm m$}\label{sec.m}

%The choice of $m$ plays a critical role in balancing robustness and estimation stability. On one hand, larger $m$ %provides more opportunities for dilution of outlier influence via aggregation; on the other hand, it reduces the sample %size per partition, potentially increasing variability. 

Recall from Theorems~\ref{thm.asymptotic_normality} and \ref{thm.IF2_phi_PCA} that the partition size $m$ does not affect the asymptotic normality of HM-PCA, but it plays an important role in its ordering-robustness. We thus propose selecting $m$ by optimizing the ordering-robustness of HM-PCA. We begin with a simple spiked covariance model as a working model, which enables an explicit characterization of HM-PCA’s ordering-robustness under contamination.

\medskip
\noindent{\bf A working model.}
Consider the following simple spiked covariance structure:
\begin{eqnarray}
\Sigma=a\xi\xi^\top+(I_p-\xi\xi^\top),\label{SSM}
\end{eqnarray}
where $\xi$ is the leading eigenvector of $\Sigma$ with signal strength $a > 1$. The target signal subspace is $\Sr = \mathrm{span}(\xi)$ with rank $r = 1$. Suppose that the data-generating distribution $F$ is contaminated by a small fraction $\varepsilon$ of samples aligned along a unit vector $\nu$ orthogonal to $\xi$ (i.e., $\nu^\top \xi = 0$), and having noise strength $\eta > 1$. Assume that all contaminated samples fall into the first block $\Xb_1$ of a random partition $[\Xb^{(m)}]$ used in HM-PCA.
Under this setting, the target of the first partition’s covariance matrix $\widehat S_1$ becomes:
\[(1 - \delta)\Sigma + \delta \eta \nu \nu^\top, \quad \text{where} \quad \delta = m \varepsilon,\]
while the remaining $\{\widehat S_k\}_{k=2}^m$ still target $\Sigma$.

\medskip
\noindent\textbf{Population-level covariance of HM-PCA.}
The population covariance matrix corresponding to HM-PCA in this contaminated setting is:
\begin{eqnarray}
\left[\frac{1}{m}\left\{(1-\delta) \Sigma + \delta \eta\nu\nu^\top\right\}^{-1}+\frac{m-1}{m}\Sigma^{-1}\right]^{-1}. \label{ex.hmpca.Sigma}
\end{eqnarray}
This matrix has the following eigenvalue-eigenvector pairs:
\begin{eqnarray*}
\left(\left\{\frac{1}{m(1-\delta)a} +\frac{m-1}{m a} \right\}^{-1},\xi\right)\quad{\rm and}\quad\left(\left\{\frac{1}{m(1-\delta +\delta\eta)}+\frac{m-1}{m}\right\}^{-1},\nu\right),
\end{eqnarray*}
%
%\left(m\left\{\frac{1}{(1-\delta)a}+\frac{m-1}{a}\right\}^{-1},\xi\right)\quad{\rm and}\quad\left(m\left\{\frac{1}{1+\delta(\eta-1)}+(m-1)\right\}^{-1},\nu\right),
%
and the remaining $(p - 2)$ eigenvalues are all equal to 
$1-\frac{\delta}{m-(m-1) \delta}$ 
%$\frac{m(1 - \delta)}{1 + (m - 1)(1 - \delta)}$ 
with eigenvectors spanning $\mathrm{span}(I_p - \xi \xi^\top - \nu \nu^\top)$.
This eigenvector $\nu$ becomes the leading eigenvector of the aggregated matrix in~\eqref{ex.hmpca.Sigma} if its associated eigenvalue exceeds that of $\xi$, namely, if 
%\begin{eqnarray*}
%\mbox{(flip condition of HM-PCA)}\quad \frac{(1-\delta)}{1+\delta(\eta-1)}<\frac{1}{a}\Big\{1-(a-1)(m-1)(1-\delta)\Big\}. 
%\end{eqnarray*}
\begin{equation}
\mbox{(flip condition of HM-PCA)}\quad 
a < \frac{[1+(m-1)(1-\delta)] (1-\delta+\delta\eta)} {(1-\delta) \left[1+(m-1)(1-\delta +\delta\eta)\right]}. 
\label{ex.hmpca_ineq}
\end{equation}
We call~\eqref{ex.hmpca_ineq} the {\it eigenvector flip condition}, since it marks the threshold at which the contaminated direction $\nu$ overtakes the signal direction $\xi$, thereby reversing their order in the eigenspectrum. 
Noting that
\[
\text{RHS of (\ref{ex.hmpca_ineq})} \;<\; \frac{1+(m-1)(1-\delta)}{(1-\delta)(m-1)}
\;=\; 1+\frac{1}{(m-1)(1-\delta)},
\]
we see that if
\begin{equation}\label{signal_strength}
a > 1+\frac{1}{(m-1)(1-\delta)},
\end{equation}
the flip condition can never be met, regardless of the value of $\eta$, even in the extreme case $\eta \to \infty$.
Exceeding the threshold in~\eqref{signal_strength} ensures immunity to contamination-induced eigenvector reordering, thereby demonstrating the optimal ordering-robustness of HM-PCA and emphasizing that its performance hinges on a proper choice of $m$.

\medskip
\noindent\textbf{Choosing $\bm m$.}  
Let $R(m)$ denote the second term in the right-hand side of~(\ref{signal_strength}), namely  
\[
R(m) = \frac{1}{(m-1)(1-\delta)}.
\]  
We choose $m$ by minimizing $R(m) + \alpha m\varepsilon$ for some penalty term $\alpha > 0$.  
The term $R(m)$ comes from the population-level signal strength threshold ensuring immunity to eigenvector reordering, and thus represents an idealized scenario without sampling variability.  
In finite samples, the estimation of eigenvalues and eigenvectors is subject to stochastic variation, typically of order $m\varepsilon$. To account for this estimation error, we introduce the penalty term $\alpha m\varepsilon$.  
Substituting $\delta = m\varepsilon$ into $R(m)$ gives $R(m) = \frac{1}{(m-1)(1-m\varepsilon)}$ and for small $\varepsilon$, the leading behavior of $R(m)$ is controlled by $\frac{1}{m-1}$.  
Minimizing $R(m) + \alpha m\varepsilon$ in this asymptotic regime yields  
$m = O(\varepsilon^{-1/2})$. 
Under the common assumption $\varepsilon = O(n^{-1})$, the optimal $m$ is of order $n^{-1/2}$, though it may not be exactly equal to $n^{-1/2}$.  
We therefore suggest the practical guideline  
\[
m = \lfloor \sqrt{n} \rfloor,
\]  
which balances the robustness gain from increasing $m$ against the loss of statistical efficiency caused by having fewer observations per subset, thereby preventing the flip condition from arising even under extremely strong contamination~$\eta$.

\section{Insights into the Ordering-Robustness of $\bm\phi$-PCA}\label{sec.compare}

\subsection{Effects of random partitioning and $\bm\phi$-aggregation}\label{sec.robust.beta1}

We have shown that $\phi$-PCA can better preserve the ordering of signal eigenvectors through the combined effects of random partitioning and $\phi$-aggregation. In this subsection, we further examine the individual contributions of these two components to ordering-robustness, as characterized by the total improvement $\tau(\cdot)$.

The contribution of random partitioning to ordering-robustness can most easily be seen by comparing AM-PCA with standard PCA. While both methods share the same underlying idea of averaging covariance matrices, AM-PCA aggregates $[\widehat S^{(m)}]$ via $\frac{1}{m}\sum_{k=1}^{m}\widehat S_k$ using separate mean estimates for each $\widehat{S}_k$, whereas standard PCA computes $\widehat{S}$ using a single global mean.  
The expression $\tau_{\rm AM}(x) = \varepsilon^2(m-1)\Delta(x)+o(\varepsilon^2)$ in Theorem~\ref{thm.robustness.3}
shows that AM-PCA still achieves an ordering-robustness gain, even though it does not make any eigenvalue transformation during aggregation. This supports the conclusion that random partitioning alone can enhance ordering-robustness.

The step of $\phi$-aggregation adds an additional layer of ordering-robustness. Consider the most severe case, where $x\in\Sr^\perp$ and has a large Mahalanobis distance $x^\top \Sigma^{-1}x$. In this setting, comparing GM-PCA with AM-PCA gives
\begin{eqnarray*}
\frac{\tau_{\text{GM}}(x)}{\tau_{\text{AM}}(x)} \approx \frac{1}{2}x^\top \Sigma^{-1}x \quad{\rm for}\quad x\in\Sr^\perp.
\end{eqnarray*}
That is, GM-PCA yields a substantially greater improvement, approximately $\frac{1}{2}x^\top \Sigma^{-1} x$ times that of AM-PCA. 
This illustrates that GM-aggregation yields a substantial ordering-robustness gain over AM-PCA when the outlier lies orthogonal to the signal subspace.
 As for HM-PCA,  it further improves the ordering-robustness gain over GM-PCA by about a factor of 2, as seen from
\begin{eqnarray}
\frac{\tau_{\text{HM}}(x)}{\tau_{\text{GM}}(x)} \approx \frac{2(x^\top \Sigma^{-1}x-1)}{x^\top \Sigma^{-1}x}\quad{\rm for}\quad x\in\Sr^\perp.\label{ratio_hg}
\end{eqnarray}
This result reinforces the earlier conclusion that HM-PCA achieves optimal ordering-robustness among the three methods.

In summary, the total improvement $\tau(x)$ in Theorem~\ref{thm.robustness.3} can be decomposed to reveal the contributions of each component:
\begin{itemize}
\item
The term $(m - 1)\Delta(x)$ represents the baseline effect of random partitioning. A larger $m$ generally yields greater ordering-robustness.

\item
The term $\frac{1-\beta}{2}x^\top \Sigma^{-1} x$ reflects the effect of $\phi$-aggregation, which grows significantly in the presence of extreme outliers. 
Specifically, GM-aggregation yields $\frac{1}{2}x^\top \Sigma^{-1} x$, while HM-aggregation further adds roughly
$\frac{1}{2}x^\top \Sigma^{-1} x$ to the total improvement, resulting in about twice the improvement over GM-aggregation. 
\end{itemize}
Note that $\phi$-aggregation is effective only when $m>1$, as $\phi$-PCA reduces to standard PCA when $m=1$, in which case the total improvement $\tau(x)$ vanishes.
These observations clarify the distinct roles of the two components: random partition provides a baseline robustness gain of approximately $\tau_{\rm AM}(x) = \varepsilon^2(m-1)\Delta(x)+o(\varepsilon^2)$, while $\phi$-aggregation delivers an additional multiplicative factor of about $\frac{1-\beta}{2}x^\top \Sigma^{-1} x$.
Together, they provide strong theoretical support for using HM-PCA in applications requiring both robustness and efficiency.

\subsection{Comparison of HM-PCA and GM-PCA}\label{sec.HM_GM}

The differences between HM-PCA and GM-PCA can be summarized in three main aspects:

\begin{itemize}
\item 
A major difference between HM-PCA and GM-PCA lies in the coefficients of the critical term $x^\top\Sigma^{-1}x$ in $\tau_{\rm HM}(x)$ and $\tau_{\rm GM}(x)$. As shown in~\eqref{ratio_hg}, HM-PCA tends to be more ordering-robust than GM-PCA.

\item 
Another difference is that $\Delta_{-1}(x)=0$ for HM-PCA, while GM-PCA can have a non-vanishing $\Delta_0(x)\ge 0$ for general $x$. A direct consequence is that the ordering-robustness gain of HM-PCA totally relies on the positivity of $\Delta(x)$. In the extreme case $\Delta(x)=0$, HM-PCA behaves like the non-robust standard PCA, while GM-PCA can still achieve a gain due to $\Delta_0(x)>0$. Thus, $\tau_{\rm GM}(x) > \tau_{\rm HM}(x)$ when $\Delta(x)\approx 0$.

\item 
HM-PCA and GM-PCA differ in their sensitivity to extremely large outliers. As discussed below equation~(\ref{ex.hmpca_ineq}), HM-PCA has the potential to completely ignore the influence of an outlier of extremely large magnitude. This property, however, does not hold for GM-PCA. Under the simple spiked model (\ref{SSM}), a similar calculation shows that the flip condition for GM-PCA is
\begin{eqnarray}
\mbox{(flip condition of GM-PCA)}\quad \eta > \frac{1-\delta}{\delta}(a^m-1).
\end{eqnarray}
This indicates that, as $\eta \to \infty$, the ordering of the signal eigenvector $\xi$ will eventually be reversed, causing GM-PCA to become biased in estimating $\Sr$. Thus, HM-PCA is generally more ordering-robust than GM-PCA in the presence of extreme outliers.
\end{itemize}

\noindent
In summary, HM-PCA and GM-PCA each have their own advantages in estimating $\Sr$. HM-PCA generally offers greater ordering-robustness than GM-PCA when $x \in \Sr^\perp$, whereas GM-PCA can outperform when $\Delta(x) = 0$. Given that $\Delta(x)$ is typically positive and that $p \gg r$ is common in practice, HM-PCA remains a compelling choice for real-world applications.

%\item
%PPCA involves the factor $\left(\frac{1}{2}x^\top\Sigma^{-1}x\right)$, while HM-PCA {\red (by setting $m=2$ for fair %comparison)} involves the factor $(1+x^\top\Sigma^{-1}x)$. This suggests that the ordering-robustness gain of HM-PCA is %larger than that of PPCA when $x\in\Sr^\perp$.

\subsection{Comparison of GM-PCA and PPCA}\label{sec.GM2_P}

Although $\phi$-PCA is motivated by generalizing the ideas of PPCA, the product aggregation $\widehat{S}_1^{\frac12} \widehat{S}_2^{\frac12}$ used in PPCA does not follow the $\phi$-PCA form, placing PPCA outside the class of $\phi$-PCA methods. Nevertheless, PPCA is closely related to GM-PCA in the case $m=2$: PPCA aggregates $\{\widehat S_1,\widehat S_2\}$ via $\widehat{S}_1^{\frac12} \widehat{S}_2^{\frac12}$, whereas GM-PCA uses
$\exp\bigl\{ \ln(\widehat{S}_1^{\frac{1}{2}}) + \ln(\widehat{S}_2^{\frac{1}{2}}) \bigr\}$
(see Remark~\ref{rmk.gm}).
Both employ a form of geometric mean for $\{\widehat S_1,\widehat S_2\}$, yet their aggregation mechanisms differ and only GM-PCA conforms to the $\phi$-PCA structure. This distinction is also reflected in Hung and Huang (2025), where the total improvement of PPCA over standard PCA, $\tau_{\rm PPCA}(x)$, admits a different approximation from $\tau_{\rm GM}(x)$, as quoted below.

\begin{prop}[Hung and Huang, 2025] \label{prop.ppca}
For any $x\in \mathbb{R}^p$ and for small $\varepsilon$, the total improvement of PPCA is expressed as
\begin{eqnarray*}
\tau_{\rm PPCA}(x)&=&
\varepsilon^2\left\{\frac{1}{2}\left(x^\top\Sigma^{-1}x\right)\Delta(x)+\Delta_{\rm PPCA}(x)\right\}+o(\varepsilon^2),
\end{eqnarray*}
where $\Delta_{\rm PPCA}(x)=\frac{p}{r(p-r)}\sum_{j\le r}\sum_{k>r}\frac{\lambda_j-\lambda_k}{2(\lambda_j+\lambda_k)}d_jd_k\ge 0$. If $x\in \Sr^\perp$, then 
$\tau_{\rm PPCA}(x)=\frac{\varepsilon^2}{2(p-r)}\left(x^\top\Sigma^{-1}x\right)^2+o(\varepsilon^2)$, which guarantees the ordering-robustness gain of PPCA.
\end{prop}

Comparing Proposition~\ref{prop.ppca} with Theorem~\ref{thm.robustness.3} yields the following observations:

\begin{itemize}
\item
For the special case $x \in \Sr^\perp$, the resemblance between GM-PCA and PPCA stems from the fact that $\tau_{\rm GM}(x)$ and $\tau_{\rm PPCA}(x)$ share the same critical term $\frac{1}{2}(x^\top\Sigma^{-1}x)$. Moreover,
\begin{eqnarray}
\frac{\tau_{\rm GM}(x)}{\tau_{\rm PPCA}(x)}&\approx& m-1\quad {\rm for}\quad x\in\Sr^\perp, \label{ratio_gm_ppca}
\end{eqnarray}
implying that, when $m = 2$, GM-PCA and PPCA are expected to perform similarly in estimating $\Sr$.

\item
For general $x$, $\tau_{\rm GM}(x)$ and $\tau_{\rm PPCA}(x)$ mainly differ in their additional terms, $\Delta_0(x)$ and $\Delta_{\rm PPCA}(x)$. Both satisfy $\Delta_0(x) \ge 0$ and $\Delta_{\rm PPCA}(x) \ge 0$ for any $x$, and both vanish when $x \in \Sr^\perp$. A straightforward calculation further shows $\Delta_{\rm PPCA}(x) \ge \Delta_0(x)$ for all $x$. These properties indicate that, while both terms enhance ordering-robustness, PPCA tends to outperform GM-PCA when $m = 2$.

\item
PPCA is developed specifically for $m = 2$, and its extension to general $m$ is not straightforward, as there is no unique definition of product-aggregation for $[\widehat S^{(m)}]$. In contrast, the aggregated covariance matrix of GM-PCA (see Remark~\ref{rmk.gm}) applies naturally to any $m$. Recall that the ordering-robustness of GM-PCA increases with $m$, as reflected by the factor $(m-1)$ in $\tau_{\rm GM}(x)$. While we expect a similar trend to hold for PPCA, extending PPCA to general $m$ and characterizing the dependence of $\tau_{\rm PPCA}(x)$ on $m$ remain open problems.
\end{itemize}

We close this section by noting that GM-PCA involves $\ln(\widehat S_k)$, which can diverge when $\widehat S_k$ is singular. To address this, we suggest implementing GM-PCA using $\ln(\widehat S_k + \epsilon I_p)$. In contrast, PPCA does not face this divergence issue, as its product-aggregation $\widehat{S}_1^{\frac12} \widehat{S}_2^{\frac12}$ is generally well-defined. This represents one advantage of PPCA, where its implementation does not require introducing the extra parameter $\epsilon$.

\section{Numerical Studies}\label{sec.simulation}

\subsection{Simulation settings}

In each simulation replicate, the signal eigenvalues are fixed as $\lambda_j=1+(p/n)^{1/2}+p^{1/(1+j)}$, $j\le r$, the noise eigenvalues $\{\lambda_j\}_{j>r}$ are generated from $U(0.5,1.5)$, and the eigenvector matrix $\Gamma$ is generated from orthogonalizing a $p\times p$ random matrix with independent $N(0,1)$ elements. Given $\Sigma=\Gamma\Lambda\Gamma^\top$, the simulation data $\Xb$ is generated from the mixture
\begin{eqnarray}
(1-\pi)N(\0_{p\times 1},\Sigma)+\pi t_1(\0_{p\times 1}, \sigma_{\rm out}^2 I_p),\label{sim.model}
\end{eqnarray}
where $t_1(\0_{p\times 1}, \sigma_{\rm out}^2I_p)$ is the multivariate $t$-distribution with 1 degree of freedom (i.e., the multivariate Cauchy distribution) having location vector $\0_{p\times 1}$ and 
scale matrix $\sigma_{\rm out}^2 I_p$. Since the Cauchy distribution has no finite moments of any order, its location and scale parameters should be interpreted in the sense of the distribution's parameterization, not as the mean or covariance.
The parameter $\pi$ is the contamination proportion. 
A larger $\sigma_{\rm out}$  tends to produce more influential outliers 
$x$ that not only deviate more substantially in direction but also yield larger values of 
$x^\top\Sigma^{-1}x$. Such outliers can dominate the leading eigenvectors of the sample covariance matrix, thereby distorting the ordering of the signal eigenvectors.

We implement HM-PCA, GM-PCA, and AM-PCA with $m=\lfloor\sqrt{n}\rfloor$ to compare with standard PCA (denoted PCA). We also implement GM-PCA with $m=2$ (denoted GM-PCA2) to assess the effect of partition size $m$ and to compare with PPCA for evaluating the GM-aggregation scheme. In addition, we implement PCA on clean data generated from $N(\0_{p\times 1},\Sigma)$ as an optimal benchmark (denoted opt-PCA). To measure the performance of each method in estimating $\Sr$, we report the similarity
\begin{eqnarray}
s_q=\frac{1}{r}\sum_{j=1}^{r}\sigma_{qj}, \quad q\ge r,
\end{eqnarray}
where $\sigma_{qj}$ is the $j$-th singular value of $B_q^\top\Gamma_r$, and $B_q$ represents the orthogonal basis of the leading $q$ eigenvectors of each method. A larger value of $s_q\in [0,1]$ indicates a better performance for $B_q$ in recovering $\Sr$, where $s_q=0$ indicates that $\Sr\perp\rmspan(B_q)$, and $s_q=1$ indicates that $\Sr\subseteq\rmspan(B_q)$. The means of $s_q$ for $q\in\{r,r+1,\ldots, 50\}$ based on 200 replicates are reported under $r=10$, $n=400$, $p\in \{200, 1000\}$, $\pi\in \{0.05, 0.1\}$, and $\sigma_{\rm out}\in\{1, 1000\}$.

%When $\sigma_{\rm out}$ is small, outliers will locate in a similar direction and will roughly perturb $\Sigma$ by one %noise eigenvector in the direction of $\mu_{\rm out}$, and hence, will have less impact on the ordering of signal %eigenvectors. 

\subsection{Simulation results}

\begin{figure}[H]
\hspace{-2.3cm}
\begin{center}
\includegraphics[width=6in]{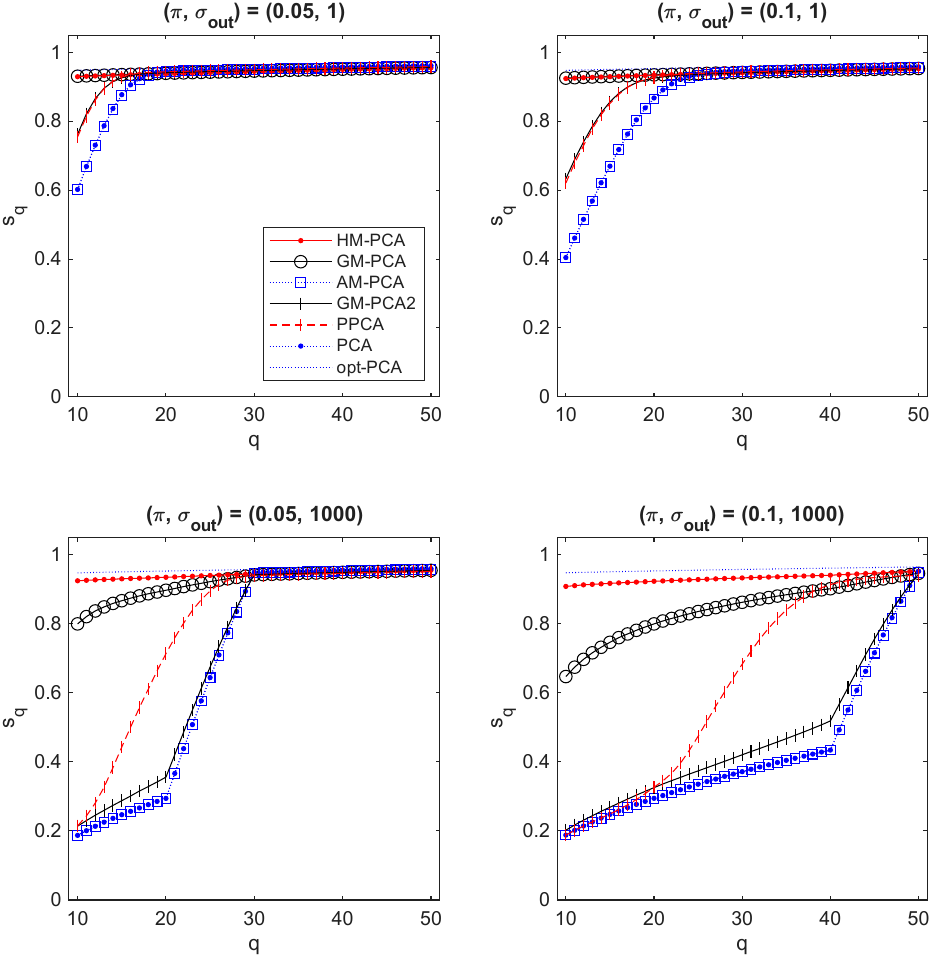}
\end{center}
\caption{The means of similarity measure $s_q$, $q\in\{r,r+1,\ldots, 50\}$ of HM-PCA, GM-PCA, AM-PCA, GM-PCA2, PPCA, PCA, and opt-PCA under $r=10$, $n=400$, $p=200$, and different combinations of $\pi\in\{0.05, 0.1\}$ and $\sigma_{\rm out}\in \{1,1000\}$.}\label{fig.sim_p200}
\end{figure}

The simulation results under $p=200$ are placed in Figure~\ref{fig.sim_p200}. We have the following observations:

\begin{itemize}
\item 
For the case of simple contamination with $(\pi,\sigma_{\rm out})=(0.05,1)$, one can see that both HM-PCA and GM-PCA 
behave similarly with opt-PCA, and they can correctly recover $\Sr$ at $q=10$ without being affected by the presence of outliers. This supports the efficiency-loss free ordering-robustness property of HM-PCA and GM-PCA. The effect of $\phi$-aggregation can be clearly seen from the differences between the $s_q$ values of HM-PCA/GM-PCA and AM-PCA, especially when $q$ is small. The effect of the partition size $m$ can be observed by comparing GM-PCA and GM-PCA2. While both methods have larger $s_q$ values than AM-PCA, GM-PCA2 is still affected by the presence of outliers, which requires more components than GM-PCA in order to recover $\Sr$. In particular, GM-PCA recovers $\Sr$ at $q=10$, while GM-PCA2 cannot recover $\Sr$ until $q=15$. AM-PCA and PCA have similar performance as expected, and they cannot well recover $\Sr$ for $q<18$. Note that AM-PCA and PCA can also achieve similar performance with opt-PCA when $q\ge 18$. This supports the fact that in the PCA problem, the estimation of signal eigenvectors is relatively robust to the presence of outliers (see also Theorem~\ref{thm.IF2_phi_PCA}, which shows that the similarities between $\gamma_j$ and $\gamma_j(F^{(m)}_{x,\varepsilon})$ share the same first- and second-order IFs). In contrast, the accurate estimation of their ordering is critical for the successful recovery of $\Sr$.

\item 
Comparing the results of $(\pi,\sigma_{\rm out})=(0.05,1)$ with $(\pi,\sigma_{\rm out})=(0.1,1)$, a lower $s_q$ curve is observed for GM-PCA2, PPCA, AM-PCA, and PCA, which indicates that the performance of these methods is adversely affected  by increasing the number of outliers. Alternatively, HM-PCA and GM-PCA are still observed to have nearly optimal performance as opt-PCA, and they can still recover $\Sr$ at the minimum size of $q=10$. This supports the good performance of $\phi$-PCA with a proper selection of $(m,\beta)$, even when the number of outliers increases.

\item 
A different behavior for HM-PCA and GM-PCA can be observed from the case of $(\pi,\sigma_{\rm out})=(0.05,1000)$. In this situation, every outlier has its unique contamination direction and can heavily disturb the estimated ordering of signal eigenvectors. In particular, there are approximately $n\pi=20$ outliers that can incorrectly dominate the leading eigenvectors. As a result, GM-PCA requires $q=r+20=30$ to achieve performance comparable to that of opt-PCA. A similar pattern is observed for the other methods, except in the case of HM-PCA. This echoes our theoretical investigation about the super robustness of HM-PCA in Section~\ref{sec.HM_GM}, where the ordering of signal eigenvectors from HM-PCA can be unchanged even when the outlier has an extremely large size. We also observe that PPCA outperforms GM-PCA2, which echoes our theoretical results in Section~\ref{sec.GM2_P} that PPCA tends to perform better than GM-PCA with $m=2$. However, GM-PCA improves GM-PCA2 and outperforms PPCA, which supports the merit of random partition with a general size $m$.

\item
The difference between HM-PCA and GM-PCA becomes most evident under the most severe contamination setting, $(\pi,\sigma_{\rm out})=(0.1,1000)$. In this case, approximately $n\pi=40$ outliers are present, and all methods, except HM-PCA, fail to recover $\Sr$ until $q=r+40=50$. Notably, HM-PCA remains unaffected by this high level of contamination and continues to achieve performance comparable to that of opt-PCA. This finding further highlights the advantage of HM-aggregation in enhancing ordering-robustness without incurring efficiency loss.
\end{itemize}

\begin{figure}[H]
\hspace{-2.3cm}
\begin{center}
\includegraphics[width=6in]{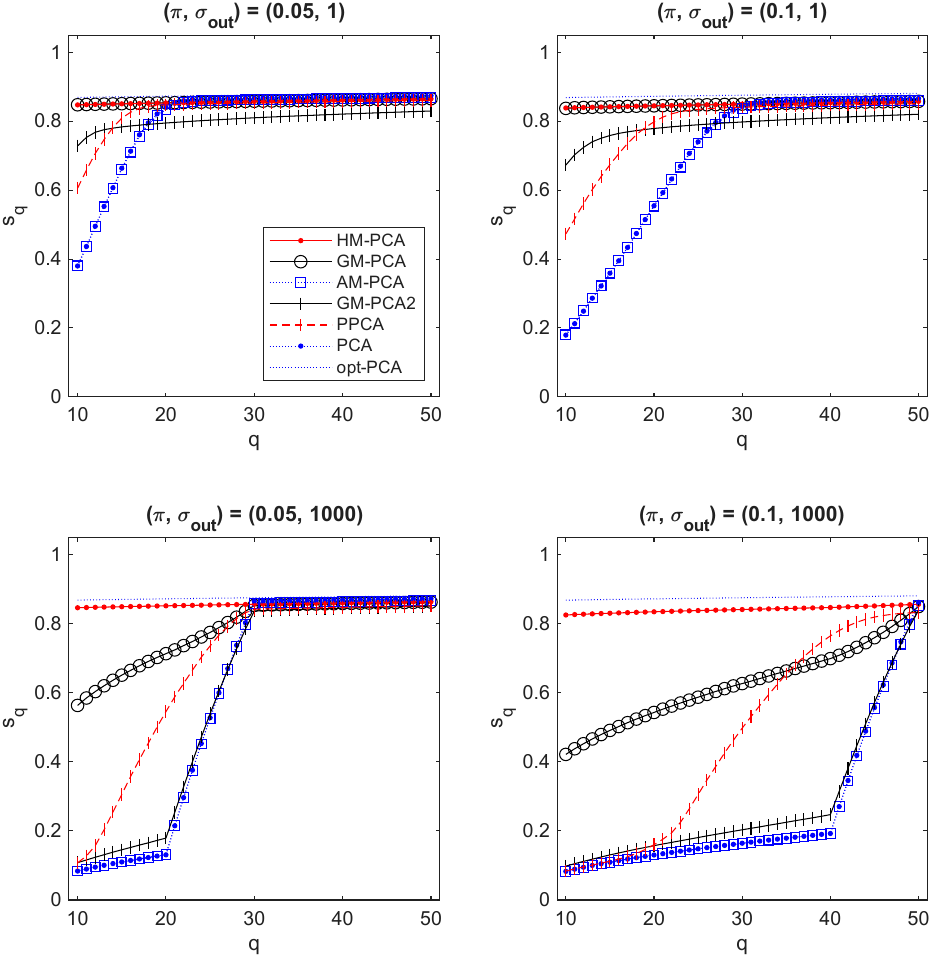}
\end{center}
\caption{The means of similarity measure $s_q$, $q\in\{r,r+1,\ldots, 50\}$ of HM-PCA, GM-PCA, AM-PCA, GM-PCA2, PPCA, PCA, and opt-PCA under $r=10$, $n=400$, $p=1000$, and different combinations of $\pi\in\{0.05, 0.1\}$ and $\sigma_{\rm out}\in \{1,1000\}$.}\label{fig.sim_p1000}
\end{figure}

Simulation results for the high-dimensional case with $p=1000$ are presented in Figure~\ref{fig.sim_p1000}. Although the efficiency-loss free ordering-robustness property of $\phi$-PCA is theoretically established under the fixed-$p$ asymptotic regime, similar conclusions to those in the case of $p=200$ can still be drawn. In particular, HM-PCA consistently outperforms all other methods, and its performance is only slightly inferior to that of opt-PCA. GM-PCA also performs well under the case of $\sigma_{\rm out}=1$, but under the case of $\sigma_{\rm out}=1000$, it fails to recover $\Sr$ until $q=50$. Moreover, the gap in $s_q$ between GM-PCA and HM-PCA becomes even more substantial than in the $p=200$ case. These findings highlight the potential sensitivity of GM-PCA to influential outliers and further support the superiority of HM-PCA in high-dimensional settings.

%all outliers tend to be around the direction of $\mu_{\rm out}$, and $\mu_{\rm out}/\|\mu_{\rm out}\|$ can be wrongly %included as a signal eigenvector. This fact is confirmed by observing that the $s_q$ values of HM-PCA, PPCA, and PCA have %a large improvement when $q$ increases from $10$ to $11$ (recall $r=10$), and they can achieve similar $s_q$ values with %opt-PCA at $q=11$. This conveys the message that the estimated ordering of signal eigenvectors are more sensitive to the %presence of outliers than their estimated directions are. 

\section{Data Analysis}\label{sec.data}

\begin{figure}[H]
\hspace{-1.5cm}\includegraphics[scale=0.3, trim=0mm 0mm 0mm 0mm, clip]{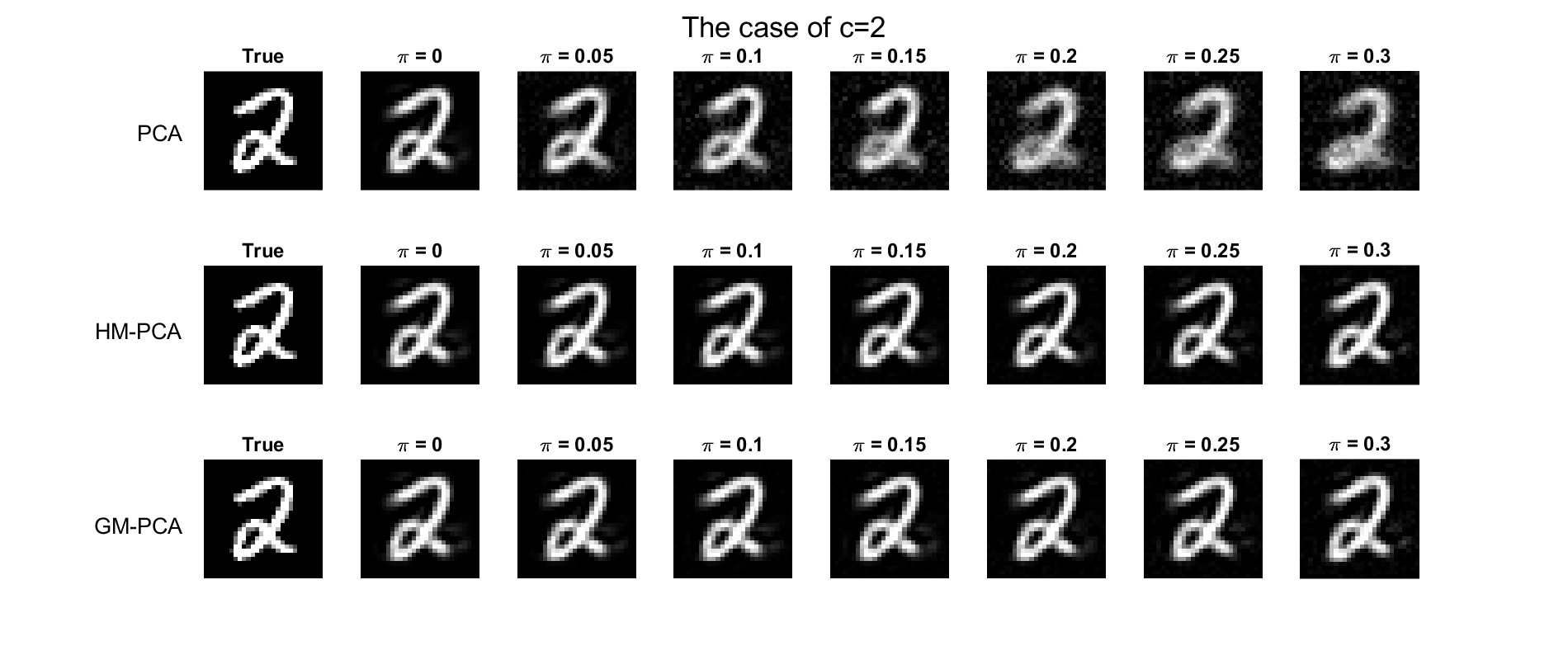}
\vspace{-2cm}
\caption{Reconstructed images of digits $1$ and $2$ with $r=50$ contamination setting (i)~$\sigma_{\rm{out}}=300$ and $\pi \in \{0, 0.05, \ldots, 0.3 \}$.}\label{fig.mnist_op}
\end{figure}

\begin{figure}[H]
\hspace{-1.5cm}\includegraphics[scale=0.3, trim=0mm 0mm 0mm 0mm, clip]{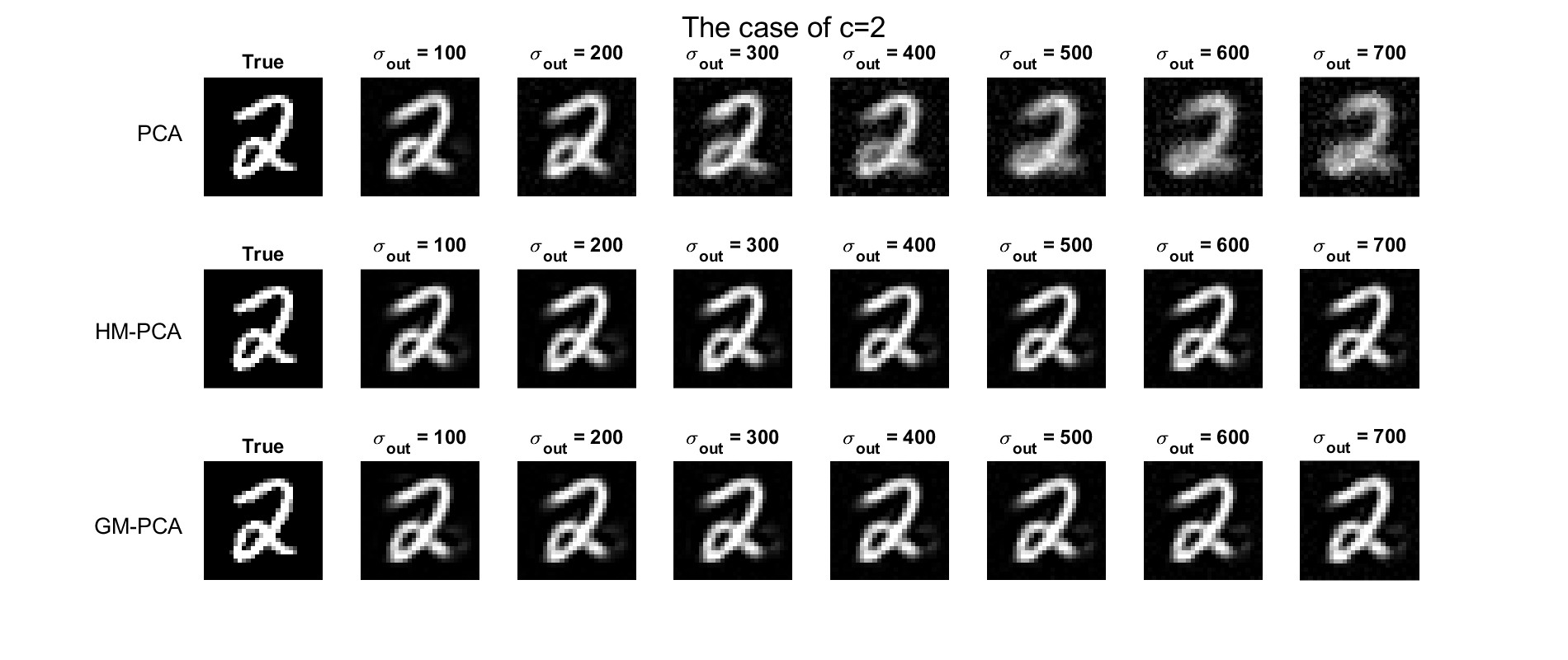}
\vspace{-2cm}
\caption{Reconstructed images of digits $1$ and $2$ with $r=50$ contamination setting (ii)~$\pi=0.1$ and $\sigma_{\rm{out}} \in \{100, 200, \ldots, 700 \}$.}\label{fig.mnist_nl}
\end{figure}

The MNIST (Modified National Institute of Standards and Technology) dataset contains $70{,}000$ grayscale images of handwritten digits, with $60{,}000$ used for training and $10{,}000$ reserved for testing. Each image has a resolution of $28 \times 28$ pixels and is labeled by its corresponding digit $c \in \{0,\ldots,9\}$. For the analysis of digit $c$, we collect the $n_c$ training images labeled as $c$, denoted by $\{X_i\}_{i=1}^{n_c}$, and construct the data matrix 
\[
\Xb = \big[ \mathrm{vec}(X_1), \ldots, \mathrm{vec}(X_{n_c}) \big]^\top \in \mathbb{R}^{n_c \times 28^2}.
\]
To evaluate robustness against contamination, we further corrupt $(100\pi)\%$ of the images in $\Xb$ by adding independent noise vectors drawn from $t_3(\mathbf{0}_{p}, \sigma_{\mathrm{out}}^2 I_p)$, where $p=28^2$.
Two contamination settings are considered:
\begin{itemize}
\item[(i)] Varying contamination proportion with fixed noise level: $\sigma_{\mathrm{out}} = 300$ and $\pi \in \{0, 0.05, \ldots, 0.3\}$.
\item[(ii)] Varying noise level with fixed contamination proportion: $\pi = 0.1$ and $\sigma_{\mathrm{out}} \in \{100, 200, \ldots, 700\}$.
\end{itemize}
Based on the data matrix $\Xb$, we apply HM-PCA, GM-PCA (with $m = \lfloor \sqrt{n_c} \rfloor$), and standard PCA to obtain the estimated subspace $\widehat{\mathcal{S}}_r$ with $r=50$. 
To evaluate performance, we reconstruct a test image $X_0$ (with label $c$) in the $r$-dimensional subspace as
\[
\mathrm{vec}(\widehat{X}_0) 
= \mathrm{vec}(\bar{X}) + P_{\widehat{\mathcal{S}}_r}\,\mathrm{vec}(X_0 - \bar{X}),
\]
where $P_{\widehat{\mathcal{S}}_r}$ denotes the projection matrix onto $\widehat{\mathcal{S}}_r$, and $\bar{X}$ is the sample mean of $\Xb$. The reconstructed vector $\mathrm{vec}(\widehat{X}_0)$ is reshaped into a $28 \times 28$ matrix to recover the image representation. Reconstruction results for digits $c \in \{1,2\}$ (with $n_1 = 6742$ and $n_2 = 5958$) under the two contamination settings are shown in Figures~\ref{fig.mnist_op}--\ref{fig.mnist_nl}.

The reconstruction results under contamination setting (i) are reported in Figure~\ref{fig.mnist_op}. When no outliers are present ($\pi=0$), HM-PCA, GM-PCA, and standard PCA produce nearly identical reconstructions, confirming that HM-PCA and GM-PCA do not incur efficiency loss compared to PCA. As the contamination proportion $\pi$ increases, however, the reconstruction quality of PCA deteriorates markedly. The images become increasingly blurred and structural details are lost. In contrast, HM-PCA and GM-PCA demonstrate substantially greater robustness, preserving the primary structure and stroke details of the digits. For example, with $c=1$, PCA already deviates noticeably from the ground truth at $\pi=0.1$, and the digit eventually disappears as $\pi$ increases. With $c=2$, PCA fails to recover the dark circular inner region of the digit’s tail, whereas HM-PCA and GM-PCA consistently maintain the original structure and produce reconstructions closer to the true image even at $\pi=0.3$.

The reconstruction results under contamination setting (ii) are reported in Figure~\ref{fig.mnist_nl}, which shows reconstructions for increasing noise levels at a fixed contamination proportion. As the noise magnitude grows, PCA reconstructions become heavily corrupted, with the digit structure and strokes progressively degraded or completely obscured. This demonstrates the sensitivity of PCA to even a small proportion of influential outliers. In contrast, HM-PCA and GM-PCA remain robust, producing stable reconstructions even at $\sigma_{\mathrm{out}} = 700$. Overall, the analysis highlights that HM-PCA and GM-PCA achieve robustness to outliers while performing comparably to standard PCA in the absence of contamination.

\section{Conclusions}\label{sec.conclusion}

In this article, we propose the $\phi$-PCA methodology as a unified framework connecting robust PCA and distributed PCA, and rigorously investigate its ordering-robustness in estimating $\Sr$. 
We derive a second-order approximation for the total improvement $\tau(x)$ of $\phi$-PCA, which disentangles the respective contributions of the aggregation function $\phi$ and the partition size $m$. 
Our analysis shows that HM-PCA, corresponding to the choice $\phi(u)=u^{-1}$, achieves optimal ordering-robustness. 
Unlike classical robust PCA procedures that mitigate outlier effects through down-weighting, HM-PCA achieves robustness via the combination of random partition and harmonic-mean aggregation. 
As a result, HM-PCA enhances ordering-robustness without sacrificing estimation efficiency relative to standard PCA.

A limitation of the present work is that the efficiency-preserving property of $\phi$-PCA is established under the fixed-$p$ regime. 
In high-dimensional settings with strong signals (i.e., $p\to\infty$ and $\lambda_r\to\infty$), Hung and Huang (2025) showed that, in the absence of outliers, PPCA and standard PCA are asymptotically equivalent in estimating $\Sr$. 
A more challenging case arises when the signals are weak (i.e., $p\to\infty$ and $\lambda_1<\infty$), where Hung, Yeh, and Huang (2025) demonstrated using random matrix theory that PPCA and standard PCA exhibit different asymptotic behaviors in both signal and noise eigenvalues. 
The tractability of PPCA in high dimensions relies on its product-aggregation structure, which admits explicit spectral analysis. 
By contrast, $\phi$-PCA involves generalized mean aggregation, to which existing techniques of random matrix theory do not directly apply. 
Extending the theoretical analysis of $\phi$-PCA to high-dimensional regimes remains an open and challenging problem, and resolving it could substantially broaden the applicability of ordering-robust methods to large-scale high-dimensional data analysis.

Finally, although the $\phi$-PCA framework is developed in the context of PCA, the underlying partition-aggregation principle—achieving robustness gains without efficiency loss—suggests a general strategy for designing robust methodologies beyond PCA.

%We also found that HM-aggregation has the potential to ignore the effect of extreme outliers, a property that is not %satisfied by other methods such as GM-aggregation.

\section*{References}
\begin{description}

\item
Arsigny, V., Fillard, P., Pennec, X., and Ayache, N. (2007). Geometric means in a novel vector space structure on symmetric positive-definite matrices. {\it SIAM Journal on Matrix Analysis and Applications}, 29(1), 328-347.

%{\red
% \item % DPCA
%Bhaskara, A. and Wijewardena, P. M. (2019). On distributed
%averaging for stochastic k-PCA. {\it Advances in Neural
%Information Processing Systems}, 32, 11024-11033.
%}
\item
Campbell, N. A. (1980). Robust procedures in multivariate analysis I: Robust covariance estimation. {\it Journal of the Royal Statistical Society, Series C (Applied Statistics)}, 29(3), 231-237.

\item
Croux, C. and Haesbroeck, G. (2000). Principal component analysis based on robust estimators of the covariance or correlation matrix: Influence functions and efficiencies. {\it Biometrika}, 87(3), 603-618.

\item % DPCA
Fan, J., Wang, D., Wang, K., and Zhu, Z. (2019). Distributed estimation of principal eigenspaces. {\it Annals of Statistics}, 47(6), 3009-3031.

\item
Fernholz, L. T. (2001). On multivariate higher order von Mises expansions. {\it Metrika}, 53(2), 123-140.

\item
Hubert, M., Rousseeuw, P. J. and Vanden Branden, K. (2005). ROBPCA: A new approach to robust principal component analysis. {\it Technometrics}, 47(1), 64-79.

\item
Hung, H. and Huang, S. Y. (2025). On the efficiency-loss free ordering-robustness of product-PCA. {\it arXiv preprint}.

\item
Hung, H., Yeh, C. C., and Huang, S. Y. (2025). On the asymptotic properties of product-PCA under the high-dimensional setting. {\it arXiv preprint}.

\item
Jolliffe, I. T. (2002). {\it Principal Component Analysis}. 2nd edition. Springer Series in Statistics. Springer, New York.

\item % beta-DPCA
Jou, Z. Y., Huang, S. Y., Hung, H., and Eguchi, S. (2025). A generalized mean approach for distributed-PCA. {\it Journal of Computational and Graphical Statistics}, 1-15. https://doi.org/10.1080/10618600.2025.2561234

%{\red
%\item % distributed SVD
%Li, X., Wang, S., Chen, K., and Zhang, Z. (2021).
%Communication-efficient distributed SVD via local power
%iterations. In {\it International Conference on Machine Learning}
%(pp. 6504-6514). PMLR.
%}

\item
Li, G. and Chen, Z. (1985). Projection-pursuit approach to robust dispersion matrices and principal components. {\it Journal of the American Statistical Association}, 80(391), 759-766.

\item
Maronna, R. A. (2005). Principal components and orthogonal regression based on robust scales. {\it Technometrics}, 47(3), 264-273.

\item  %  DPCA review
Wu, S. X., Wai, H. T., Li, L., and Scaglione, A. (2018). A review of distributed algorithms for principal component analysis. {\it Proceedings of the IEEE}, 106(8), 1321-1340.

\item
Yata, K. and Aoshima, M. (2010). Effective PCA for high-dimension, low-sample-size data with singular value decomposition of cross data matrix. {\it Journal of Multivariate Analysis}, 101(9), 2060-2077.
\end{description}

\clearpage

\clearpage

\begin{figure}[H]
\hspace{-1.5cm}\includegraphics[scale=0.3]{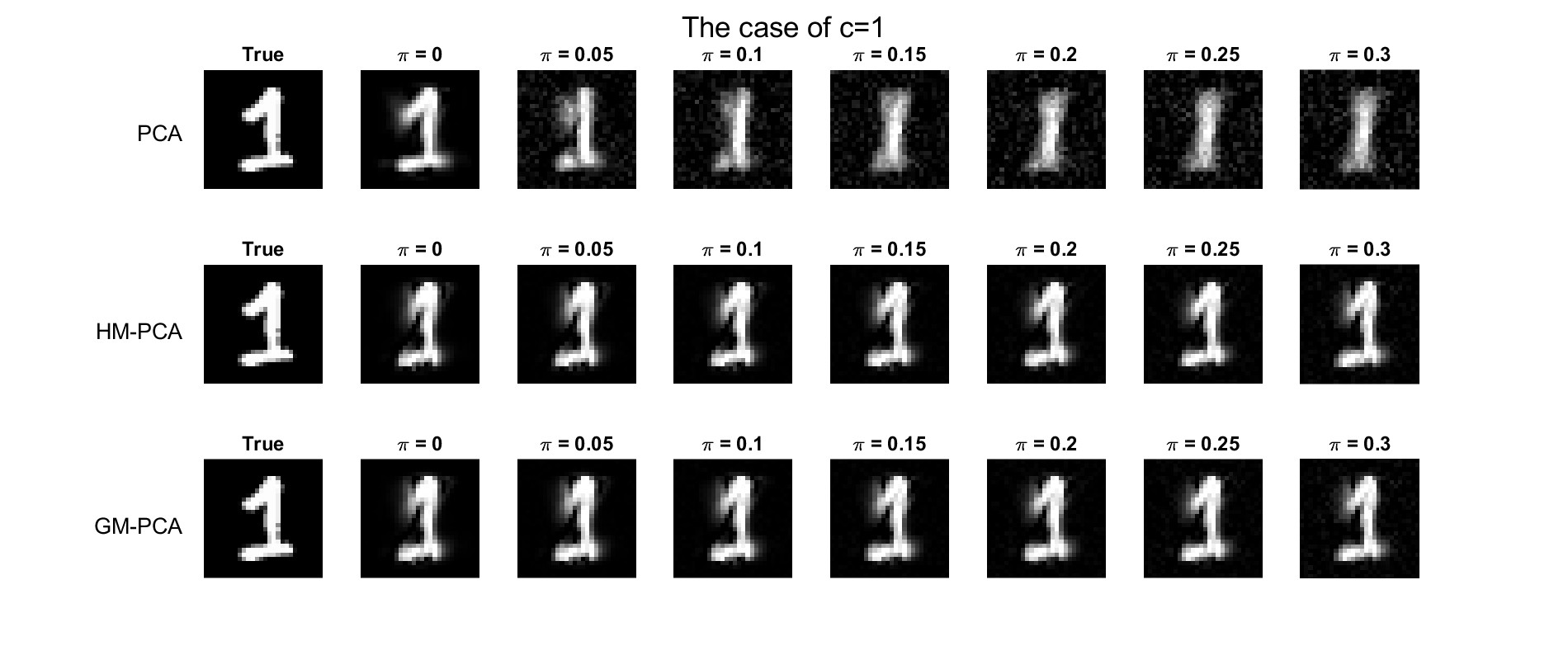}
\vspace{-1.8cm}
\end{figure}

\begin{figure}[H]
\hspace{-1.5cm}\includegraphics[scale=0.3]{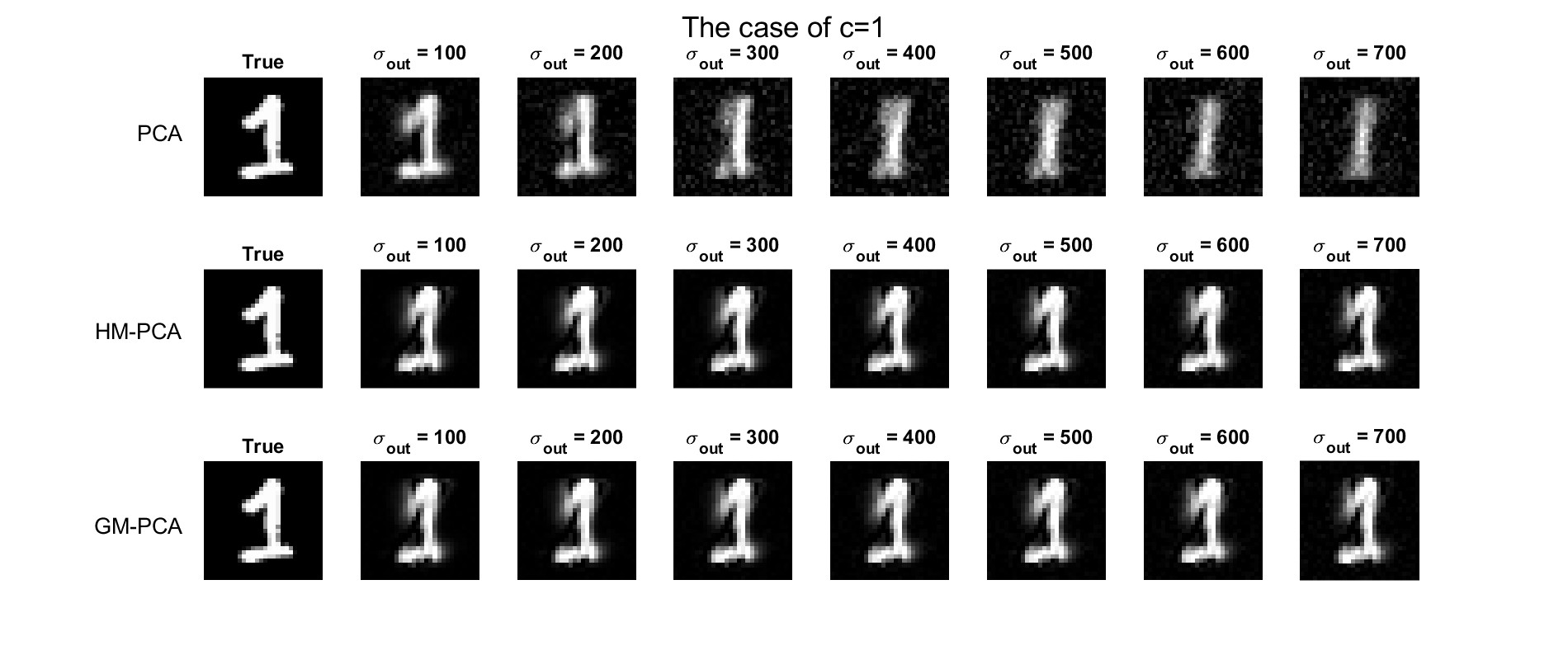}
\vspace{-1.8cm}
\end{figure}

\end{document}